\pdfoutput=1
\documentclass[apjl]{emulateapj}
\usepackage{color}
\usepackage{amsmath}
\usepackage{ulem}

\usepackage{txfonts} % looks nicer

\definecolor{brown}{rgb}{0.42,0.24,0.07}
\definecolor{darkgreen}{rgb}{0.0,0.6,0.00}
\definecolor{purple}{rgb}{0.7,0.0,0.7}
\definecolor{black}{rgb}{0.0,0.0,0.0}

\newcommand{\Sigmagas}{\ensuremath{\varSigma_\mathrm{g}}}
\newcommand{\Sigmadust}{\ensuremath{\varSigma_\mathrm{d}}}
\newcommand{\Sigmagasref}{\ensuremath{\varSigma_\mathrm{g,0}}}
\newcommand{\OmegaK}{\ensuremath{\varOmega_\mathrm{K}}}
\newcommand{\Omegagas}{\ensuremath{\varOmega}}

\newcommand{\rhopmat}{\ensuremath{\rho_\mathrm{mat}}}
\newcommand{\epsdtog}{\ensuremath{\epsilon_\mathrm{d/g}}}
\newcommand{\phieffi}{\varPhi_{\mathrm{eff}}}

\newcommand{\tauf}{\tau_\mathrm{f}}

\newcommand{\alphaSS}{\ensuremath{\alpha}} % Shakura-Sunyaev alpha
\newcommand{\alphaSSref}{\ensuremath{\alpha}} % Shakura-Sunyaev alpha at reference radius
\newcommand{\tauchar}{\ensuremath{\tilde{\tau}}} % optical depth
\newcommand{\betaPR}{\ensuremath{\beta_\mathrm{PR}}} % Poynting-Robertson beta
\newcommand{\rBR}{\ensuremath{r_\mathrm{BR}}} % reference radius (birth ring)

\newcommand{\ThetaPE}{\ensuremath{\Theta}}
\newcommand{\St}{\ensuremath{\mathrm{St}}}

\newcommand{\Mstar}{\ensuremath{M_\star}}
\newcommand{\Lstar}{\ensuremath{L_\star}}
\newcommand{\dragcoef}{\ensuremath{C_D}}
\newcommand{\e}{\ensuremath{e}}
\newcommand{\grainradius}{\ensuremath{a}}
\newcommand{\cs}{\ensuremath{c_s}}
\newcommand{\vv}{\ensuremath{\v{v}}}
\newcommand{\Msun}{\ensuremath{M_\odot}}
\newcommand{\Mearth}{\ensuremath{M_\oplus}}
\newcommand{\Mdust}{\ensuremath{M_\mathrm{dust}}}
\newcommand{\AU}{\ensuremath{\mathrm{AU}}}
\newcommand{\micrometer}{\ensuremath{\muup\mathrm{m}}}
\newcommand{\Lsun}{\ensuremath{L_\odot}}
\newcommand{\cm}{\ensuremath{\mathrm{cm}}}
\newcommand{\gram}{\ensuremath{\mathrm{g}}}
\newcommand{\gcmtwo}{\ensuremath{\gram~\cm^{-2}}}
\newcommand{\deriv}[2]{\frac{d{#1}}{d{#2}}}
\renewcommand{\v}[1]{{\boldsymbol{#1}}}
\newcommand{\aderiv}[1]{\frac{D{#1}}{D t}}
\newcommand{\del}{\v{\nabla}}
\newcommand{\grad}{\del}
\newcommand{\Div}{\del\cdot}
\newcommand{\runcaption}[3]{Gas (left panel) and dust (right panel) surface
densities after 400 orbits for run #1
($\Sigmagasref=#2$; $\Mdust=#3$). Dashed circle indicates {\rBR} (100~AU).}

\newcommand{\gasA}{\ensuremath{1.2\times10^{-7}}}
\newcommand{\gasB}{\ensuremath{1.2\times10^{-6}}}
\newcommand{\gasC}{\ensuremath{1.2\times10^{-5}}}
\newcommand{\gasD}{\ensuremath{1.2\times10^{-4}}}
\newcommand{\gasE}{\ensuremath{1.2\times10^{-3}}}
\newcommand{\gasF}{\ensuremath{1.2\times10^{-3}}}
\newcommand{\gasG}{\ensuremath{1.2\times10^{-3}}}
\newcommand{\dustA}{\ensuremath{9\times10^{-3}}}
\newcommand{\dustB}{\ensuremath{9\times10^{-3}}}
\newcommand{\dustC}{\ensuremath{9\times10^{-3}}}
\newcommand{\dustD}{\ensuremath{9\times10^{-3}}}
\newcommand{\dustE}{\ensuremath{9\times10^{-3}}}
\newcommand{\dustF}{\ensuremath{9\times10^{-2}}}
\newcommand{\dustG}{\ensuremath{9\times10^{-2}}}

\newcommand{\Fig}[1]{Fig.~\ref{#1}}
\newcommand{\fig}[1]{\Fig{#1}}

\newcommand{\Eq}[1]{Eq. (\ref{#1})}

\newcommand{\eq}[1]{\Eq{#1}}

\slugcomment{Submitted to ApJ, September 12, 2017}

\shorttitle{Radiation pressure and the photoelectric instability}
\shortauthors{Richert et al.}

\begin{document}

\title{The interplay between radiation pressure and the photoelectric instability in optically thin disks of gas and dust}
\author{Alexander J.W. Richert\altaffilmark{1,2}, Wladimir Lyra\altaffilmark{3,4,5}, \& Marc Kuchner\altaffilmark{2}}
\altaffiltext{1}{Department of Astronomy \& Astrophysics, Penn State University, 525 Davey Lab, University Park, PA 16802. ajr327@psu.edu} 
\altaffiltext{2}{Exoplanets and Stellar Astrophysics Laboratory, NASA Goddard Space Flight Center, 8800 Greenbelt Road, Greenbelt, MD 20771.}
\altaffiltext{3}{Department of Physics and Astronomy, California State University Northridge, 18111 Nordhoff St, Northridge, CA 91330.}
\altaffiltext{4}{Jet Propulsion Laboratory, California Institute of Technology, 4800 Oak Grove Drive, Pasadena, CA 91109.}
\altaffiltext{5}{Division of Geological \& Planetary Sciences, California Institute of Technology, 1200 E California Blvd MC 150-21, Pasadena, CA 91125.}

\begin{abstract}

Previous theoretical works have shown that in optically thin disks, dust grains
are photoelectrically stripped of electrons by starlight, heating nearby gas
and possibly creating a dust clumping instability---the photoelectric
instability (PeI)---that significantly alters global disk structure. In the
current work, we use the Pencil Code to perform the first numerical models of
the PeI that include stellar radiation pressure on dust grains in order to
explore the parameter regime in which the instability operates. In models with
{gas surface densities greater than $\sim$$10^{-4}~\gram~\cm^{-2}$}, we see a variety of dust structures,
including sharp concentric rings and{, for models with especially high gas surface densities ($\sim$$10^{-3}~\gram~\cm^{-2}$),} non-axisymmetric arcs and clumps that
represent dust surface density enhancements of factors of $\sim$5--20 depending
on the run parameters. The gas distributions show various structures as well,
including clumps and arcs formed from spiral arms. In models with lower gas
surface densities, vortices and smooth spiral arms form in the gas distribution,
but the dust is too weakly coupled to the gas to be significantly perturbed. In
one high gas surface density model, we include a large, low-order gas viscosity,
and, in agreement with previous radiation pressure-free models, find that it
observably smooths the structures that form in the gas and dust, suggesting that
resolved images of a given disk may be useful for deriving constraints on the
effective viscosity of its gas. Broadly, our models show that radiation pressure
does not preclude the formation of complex structure from the PeI, but the
qualitative manifestation of the PeI depends strongly on the parameters of the
system. The PeI may provide an explanation for unusual disk morphologies such as
the moving blobs of the AU~Mic disk, the asymmetric dust distribution of the
49~Ceti disk, and the rings and arcs found in the disk around HD~141569A.

\end{abstract}

\section{Introduction}
\label{s:introduction}

Circumstellar disks play a key role in testing theories of planet formation and
evolution, revealing the physical and chemical environment of planet-forming
systems, including providing constraints on the properties of nascent planets.
{Resolved images of protoplanetary disks, transitional disks, and debris disks
show a variety of complex morphologies, including cavities, gaps and rings \citep{Debes2013, Wahhaj2014, Follette2015, Currie2015, vanBoekel2017}, as well crescent-shaped structures, arcs, and spiral arms \citep{vanderMarel2013, Grady2013, Biller2015, Perrot2016, Follette2017}. These disk structures are frequently attributed to gravitational
perturbation by unseen embedded planets \citep[e.g.,][]{KuchnerHolman2003, NesvoldKuchner2015, Richert2015, Dong2016, Dipierro2017, Dong2017a, Dong2017b}.}

The possibility of comparable masses of gas and dust in any given optically
thin disk raises the possibility of hydrodynamical interactions that will give
rise to features like gaps, rings, and clumps that are frequently attributed to
gravitational perturbation by an unseen embedded planet \citep{Klahr2005,
Besla2007, LyraKuchner13}. In optically thin disks, stellar far ultraviolet
photons whose energies exceed the work function of the dust grains \citep[a few
eV;][]{Besla2007} photoelectrically eject electrons which in turn heat nearby
gas. \citet{Klahr2005} and \citet{Besla2007} propose that this leads to a
clumping instability---the photoelectric instability (PeI)---wherein heating of
the gas by dust grains creates a local pressure maximum, which then traps more
dust, which further heats the gas, and so on. \citet{Klahr2005} model the
system in 1D, finding the instability, and extrapolate from the 1D results to
suggest that in 2D the photoelectric instability will generate ring structures,
similar to those observed in disks like the one around HR~4796A.
\citet{LyraKuchner13} model the system hydrodynamically with 2D global and 3D
local simulations, and find that rings are not formed unless the backreaction
of the drag force is considered. When that component is ignored, power
concentrates in high azimuthal wavenumbers and only clumps are formed. When the
action of the dust on the gas is considered, rings and incomplete arcs are seen
to form in the dust distribution.

Other findings of \citet{LyraKuchner13} are that 1) linear instability exists only
for dust-to-gas ratio $\epsdtog < 1$, with maximum growth rate at $\epsdtog
\approx 0.2$; 2) non-linear instability is observed for $\epsdtog=1$; 3) linear
instability only exists if photoelectric heating is the dominant heating source;
4) the photoelectric instability supersedes the streaming instability when the
conditions for both are present; and 5) the particular mode for which gas and
dust velocity are equal, thus canceling the drag force and backreaction, executes
free oscillations, which are seen as a small but finite eccentricity
($\approx$0.03).

The photoelectric instability may provide an explanation for a number of
observed systems with unusual morphologies. Scattered light images of the AU~Mic
disk, an edge-on system, reveal radially moving blobs not seen at longer
wavelengths. The disk around HIP~73145 contains concentric rings in scattered
light images (which reveal small grains), while larger grains, observed at ALMA
wavelengths, are distributed more compactly around the central star
\citep{Feldt2017}. For both of these systems, the differing behavior of small
and large grains is not readily explained by a planetary perturber. The edge-on
disk around 49~Ceti is known to be gas rich, though the total mass remains
poorly constrained \citep{Hughes2017}; \citet{Hughes2017} identify asymmetric
structure in the disk consistent with a warp or spiral arm, finding no such
features in two resolved gas-poor disks. The HD~141569A transition disk contains
rings and arclets of small grains \citep{Perrot2016}. The findings of
\citet{Klahr2005}, \citet{Besla2007}, and \citet{LyraKuchner13} also raise the
question of whether the presence of gas plays a role in the formation of the
sharp dust rings seen in scattered light images of disks such as those around
HR~4796A \citep{Milli2017} and Fomalhaut \citep{Kalas2008}.

As novel as they are, the hydrodynamical models of \citet{LyraKuchner13} do not
include radiation pressure from the central star on dust grains, which even
around low-mass stars will put sufficiently small grains on highly eccentric
orbits, in some cases blowing them out of the system. The ability of the
photoelectric instability to explain the morphologies of optically thin disks
depends vitally on whether it can operate in the presence of stellar radiation
pressure on dust grains.

In this work, we conduct hydrodynamical simulations of optically thin disks that
include both dust--gas photoelectric heating and radiation pressure on dust
grains that span a range of sizes. In Section~\ref{s:grainsize}, we provide an
analytical discussion of the role grain size with respect to the emergence of
hydrodynamical instabilities. Equations solved and initial conditions are
discussed in Section~\ref{s:methods}. Results are discussed in
Section~\ref{s:results}, while further conclusions and implications for future
work are discussed in Section~\ref{s:conclusions}.

\section{The role of grain size} \label{s:grainsize}

A priori, it may be expected that radiation pressure from the central star on
dust grains in an optically thin disk will inhibit the formation of the clumps,
arcs, and rings seen in the models of \citet{LyraKuchner13}. For spherical grains,
the radiation pressure strength $\beta$, defined as the ratio of the radiation
force to the gravitational force, depends on host star mass {\Mstar}, host star
luminosity {\Lstar}, grain radius {\grainradius}, and dust material density
{\rhopmat} \citep{Burns1979, Krivov2010}, such that
\begin{equation} \label{e:beta}
\beta \simeq 0.574 \ \frac{\Mstar}{\Msun} \ \left(\frac{\Lstar}{\Lsun}\right)^{-1} \ \left(\frac{\grainradius}{1~\micrometer}\right)^{-1} \ \left(\frac{\rhopmat}{1~\gram\,\cm^{-3}}\right)^{-1} .
\end{equation}
When a dust grain is created on a Keplerian orbit, radiation pressure places it
on an eccentric orbit where eccentricity $\e = \beta/(1-\beta)$
\citep{Burns1979, Strubbe2006}. A grain with $\beta=\frac{1}{2}$ receives a
radiation force equal to half the gravitational force, causing it to become
unbound ($\e = 1$). The models produced in this paper will help to determine
whether non-zero orbital eccentricities of dust grains affect the onset of
clumping instabilities.

Clumping due to the photoelectric instability depends on aerodynamic drag.
\citet{LyraKuchner13} find that the instability is robust for this variable, meaning
that grains of longer stopping time simply take longer to respond to the
pressure maximum and concentrate. Yet, one can imagine that if other dynamical
processes are modifying the state of the gas at timescales shorter than the
stopping time of the grains, clumping by the photoelectric instability may be
disrupted. In other words, although linear growth is present, the saturated
state may be quite different for small and big grains. Given the stopping time
$\tauf$, a nondimensional stopping time can be constructed, also known as
Stokes number, $\St\equiv\tauf\OmegaK$, where $\OmegaK$ is the
Keplerian frequency. For a thin disk,
\begin{equation} \label{e:dragtime}
{\St} \approx \frac{\pi \grainradius \rhopmat}{2 \Sigmagas} ,
\end{equation}
where {\Sigmagas} is the surface density of the gas. Grain size therefore seems
likely to play a dual role in the development of dust clumping instabilities in
optically thin disks, which require grains small enough to be susceptible to
aerodynamic drag, but large enough to remain bound to the star, preferably on a
low-eccentricity orbit.

\begin{figure}
\resizebox{\columnwidth}{!}{\includegraphics{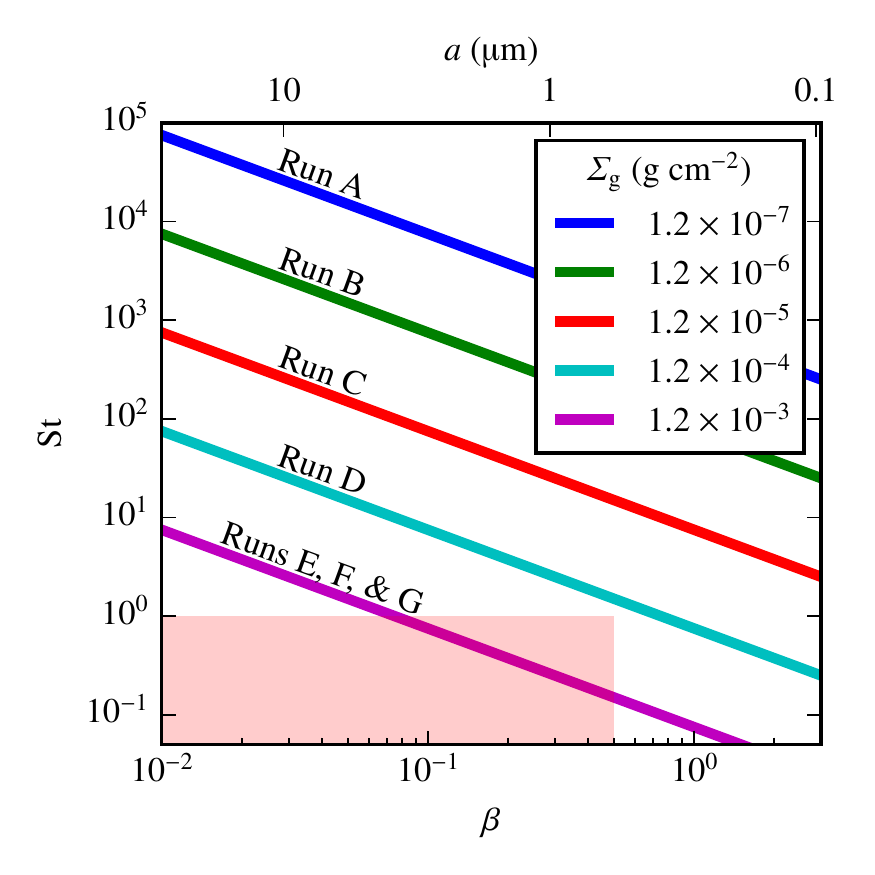}}
\caption[{\St} vs. $\beta$ for a solar-type star]{Dimensionless drag stopping time {\St} as a function of
$\beta$ for several values of gas surface density {\Sigmagas}, assuming a
solar-type star. Pink shaded area indicates bound grains
($\beta<\frac{1}{2}$) with short stopping times
($\St<1$).}
\label{f:tauvsbeta}
\end{figure}

The relationship between {\St} and $\beta$ is easily specified for
spherical grains in a thin disk based on Equations~\ref{e:beta} and \ref{e:dragtime},
such that
\begin{equation} \label{e:tauvsbeta}
{\St} = \frac{1}{\beta} \ \left(\frac {\Sigmagas}{9\times10^{-5}~{\gram}\,{\cm}^{-2}}\right)^{-1} \ \frac{\Lstar}{\Lsun} \ \left(\frac {\Mstar}{\Msun}\right)^{-1} .
\end{equation}
In \fig{f:tauvsbeta}, we show {\St} as a function of $\beta$ for
a solar-type star for several values of {\Sigmagas}. The gas surface density
values shown are a few orders of magnitude below the densities where the optical
thickness of the gas will impede both photoelectric stripping of dust and
radiation pressure.

A priori, it would seem that a substantial level of gas is required in order to
rapidly create dust clumping instabilities in the presence of radiation
pressure, while larger grains may be able to generate such instabilities over
longer timescales \citep{LyraKuchner13}. Yet, clumping by PeI in the presence of
radiation pressure may require grains of Stokes number near unity. A shorter PeI
onset timescale would make the PeI to operate in disks spanning a large range of
dust production rates (which remain poorly constrained in observed systems).

\section{Methods}
\label{s:methods}

\subsection{Equations}
\label{ss:equations}

In this work we perform two-dimensional global simulations of gas-bearing
optically thin circumstellar disks using the Pencil Code, a high-order finite
difference hydrodynamics code \citep{Brandenburg2002}. Both the gas and dust
are calculated in Cartesian coordinates\footnote{In the course of this study,
we identified a bug in the Pencil Code that affected the calculation of
azimuthal particle accelerations in polar coordinates. The bug has since been
fixed, and does not seem to have adversely affected previous works using that
part of the code.}. The code evolves the gas according to the continuity
equation,
\begin{equation}\label{e:continuity}
  \aderiv{\Sigmagas} = -\Sigmagas \Div \v{u},
\end{equation}
and the equation of motion,
\begin{equation}\label{e:motion}
 \aderiv{\v{u}} = -\frac{1}{\Sigmagas}\grad P - \grad \varPhi_\star - \frac{\Sigmadust}{\Sigmagas} f_d,
\end{equation}
where {\Sigmagas} and {\Sigmadust} are the gas and dust surface densities, and
$\v{u}$ and $P$ are the velocity and pressure of the gas. $\varPhi_\star$
represents the Newtonian gravitational potential of the star. The dust drag
term $f_d$ is defined below.

We model the dust using 400,000 Lagrangian superparticles of equal mass, except
for two runs where the total dust mass is increased in part by doubling the
number of superparticles to 800,000 (see \S~\ref{ss:modelparameters}). Each
superparticle contains subparticles of one physical radius {\grainradius}
between 0.1~{\micrometer} and 10~{\micrometer}, and all subparticles have a
constant material density $\rhopmat = 2~\rm{g}\,\rm{cm}^{-3}$, the approximate
material density of silicate grains. This yields a total dust mass of
$\sim0.01~\Mearth$.

The overall grain size distribution follows the standard \citet{Dohnanyi1969}
$q=-3.5$ power law. The use of superparticles of the same mass automatically
contributes a dependence with $q=-3$ (superparticles with smaller grains will
contain more particles); grain sizes associated with superparticles follow a
$q=-0.5$ dependence in order to yield a $q=-3.5$ size distribution overall. This
scheme prevents numerical issues associated with consolidating large grains into
a small numbers of very massive superparticles (which, especially when modeling
photoelectric heating, can crash the code).

In all simulations, superparticles are inserted in a birth ring positioned at
100~AU from the central star; the birth ring is axisymmetric and has a radial
Gaussian profile ($\sigma=10~\AU$) in order to avoid an artificial sharp edge in
the dust distribution. Superparticles are at first inserted gradually over the
course of several orbits in order to avoid discontinuities and ensure that
superparticles span a range of orbital phases, but then are inserted throughout
the simulation so as to yield a constant number of superparticles, and therefore
constant total mass of the disk. That is to say, when a superparticle crosses
the inner (50~AU) or outer (800~AU) boundary of the disk, another superparticle
is inserted in the birth ring with a new grain size chosen at random according
to the $q=-3.5$ distribution.

The dynamical equation for each dust superparticle with velocity $\vv$ depends
on the effective gravitational potential $\phieffi$ and gas drag term
$f_\mathrm{d}$,
\begin{equation}\label{e:dust}
  \deriv{\vv}{t} = -\phieffi + f_\mathrm{d} .
\end{equation}
The expression for the drag acceleration is given by
\begin{equation}\label{e:drag}
  f_d = -\left( \frac{2 \OmegaK \Sigmagas \dragcoef}{\pi \grainradius \rhopmat} \right) \Delta \vv ,
\end{equation}
where $\OmegaK$ is the Keplerian orbital frequency at a given orbital radius, and
gas--dust velocity differential $\Delta \vv = \vv - \v{u}$. The low gas
densities of debris disks imply very large mean free paths compared with dust
grain radii (i.e., $\lambda \gg \grainradius$), hence the drag coefficient {\dragcoef} is
calculated for the Epstein regime as a function of sound speed {\cs} such that
\begin{equation}\label{e:epsteincoef}
  C_D = \sqrt{1+\frac{9\pi}{128} ( |\Delta \vv|/\cs )^2} .
\end{equation}
The effective gravitational potential term $\phieffi$ incorporates radiation
pressure on the dust, such that
\begin{equation}\label{e:rp}
  \phieffi = \frac{G M_\star(1-\beta)}{r^2} ,
\end{equation}
where for each superparticle, the ratio of the radiation pressure force to
gravitational force $\beta = \beta_{\rm ref}/a$. We calculate reference
radiation pressure strength $\beta_{\rm ref}$ for a solar-type star according to
the prescription of \citet{Burns1979}, such that $\beta \approx 0.2$ for a
1~{\micrometer} grain with density 2~g\,cm$^{-3}$. Our grain size range
corresponds with radiation pressure strengths $0.03<\beta<3$, which in turn
corresponds with eccentricities ranging from near-circular to completely unbound
orbits, where an unbound orbit ($e \geq 1$) has $\beta>0.5$, corresponding with
grain size $\grainradius<0.57~\micrometer$.

Given that the photoelectric heating time is small compared with the dynamical
time \citep{Besla2007, LyraKuchner13}, we adopt the modified equation of state of
\citet{LyraKuchner13} that implements instantaneous heating of gas by dust. The gas
pressure $P$ is assumed to be proportional to the dust density {\Sigmadust},
such that
\begin{equation}
\grad P = \frac{\ThetaPE c^2_{s0}}{\gamma \Sigmagasref} (\Sigmagas \grad \Sigmadust + \Sigmadust \grad \Sigmagas) .
\end{equation}
{\ThetaPE} is a dimensionless parameter that sets the pressure contribution of
photoelectric heating compared to the background temperature of the gas \citep{LyraKuchner13},
which is itself specified by the reference sound speed $c_{s0} = 0.05$ in code
units, corresponding with a scale height of 0.05 which is assumed for both the
gas and dust. The value of {\Sigmadust} is calculated by interpolating particle
positions onto the grid using a triangular-shaped cloud particle mesh scheme
\citep{Eastwood1974}. We assume that the gas is locally isothermal, which is
appropriate for the short cooling times expected in debris disks
\citep{LyraKuchner13}.

The high-order scheme used by the Pencil Code leads to little numerical
dissipation, therefore we apply sixth-order hyperdissipation terms to the
r.h.s. of Equations \ref{e:continuity} and \ref{e:motion} to stabilize the
density and velocity fields, respectively, at the grid scale
\citep{Lyra2008, McNally2012, Lyra2017}.

\subsection{Model parameters}
\label{ss:modelparameters}

We conduct seven simulations in total. For five models---runs A--E---we vary
the initial gas surface density {\Sigmagasref} according to the values shown in
\fig{f:tauvsbeta}. This range of surfaces densities corresponds with surface
densities $10^{4}-10^{8}$ times lower than that of the Minimum Mass Solar
Nebula at 100~AU \citep{Weidenschilling1977}. For the $\beta$~Pic disk,
\citet{Brandeker2004} assume solar abundances to derive a gas surface density
of $3.5 \times 10^{-6}~\gcmtwo$, based on a scale height of 10~AU and mean
molecular mass 2.5~amu. Our models have gas surface densities approximately
0.03 to 300 times that value. In run F, we duplicate our model with the highest
value of {\Sigmagasref} (run E) but increase the total dust mass by a factor of
10; we achieve this by doubling the number of superparticles and increasing the
mass per superparticle by a factor of five. In run G, we duplicate run F but
add a large Laplacian viscosity to the r.h.s. of \eq{e:motion}, corresponding
with a Shakura--Sunyaev {\alphaSS} viscosity of {0.1} \citep{Shakura1973} at
the reference orbital radius (100~AU). \citet{LyraKuchner13} find that
viscosity damps the PeI at high wavenumbers. This viscosity is presumably
expected from the magnetorotational instability \citep[MRI;][]{BalbusHawley91}.
Although the activity of the MRI in optically thin disks remains limitedly
understood \citep{Kral2016mri}, we include this additional run to explore the
effect of an eddy viscosity when photoelectric heating and radiation pressure
are both operative.

The physical parameters that differentiate our models are summarized in
Table~\ref{t:runs}, including the Stokes number of the smallest particle in
each run, min({\St}) (calculated using Eq.~\ref{e:dragtime} for
$\Sigmagas=\Sigmagasref$). Table~\ref{t:runs} also provides the spatially
averaged dust-to-gas ratio {\epsdtog} for $r<300~\AU$ after 400 orbits, denoted
as $\langle\epsdtog\rangle$, as well as the characteristic vertical geometric
optical depth of any dust overdensities after 400 orbits, denoted as
{\tauchar}.

\begin{table*}
\caption{Run parameters.}
\begin{center}
\begin{tabular}{ccccccc}
\hline
Run     &   {\Sigmagasref}       &    Total dust mass       &    {\alphaSSref} & min(\St)            &   $\langle\epsdtog\rangle$  & {\tauchar}            \\ 
        &   (\gram~\cm$^{-2}$)   &    (\Mearth)             &                  &                     &   (400 orbits)                                        \\ 
\hline                                                                           
A       &   \gasA                &    \dustA                &    ---           & $2.6\times10^{2}$   &   5                         & $7\times10^{-3}$       \\ 
B       &   \gasB                &    \dustB                &    ---           & $2.6\times10^{1}$   &   0.8                       & $6\times10^{-3}$       \\ 
C       &   \gasC                &    \dustC                &    ---           & $2.6\times10^{0}$   &   0.1                       & $4\times10^{-3}$       \\ 
D       &   \gasD                &    \dustD                &    ---           & $2.6\times10^{-1}$  &   0.003                     & $5\times10^{-4}$       \\ 
E       &   \gasE                &    \dustE                &    ---           & $2.6\times10^{-2}$  &   0.0001                    & $4\times10^{-4}$       \\ 
F       &   \gasF                &    \dustF                &    ---           & $2.6\times10^{-2}$  &   0.006                     & $1\times10^{-2}$       \\ 
G       &   \gasG                &    \dustG                &    0.1           & $2.6\times10^{-2}$  &   0.002                     & $1\times10^{-2}$       \\ 
\hline
\end{tabular}
\end{center}
\label{t:runs}
\end{table*}

In each model, the gas is assumed to be initially uniformly distributed
throughout the disk between 50~AU and 800~AU from the central star. The gas
temperature is also uniform throughout the disk. This yields no global pressure
gradient, allowing us to isolate the effects of photoelectric heating;
specifically, it allows us to attribute any radial dust drift to the
radiation pressure and PeI. \citet{LyraKuchner13} show for the radiation pressure-free case that the photoelectric instability generates dust rings in the presence of a global pressure gradient and the streaming instability.

We run each model for 400 orbits, a sufficient amount of time to determine
whether small-to-medium grains can trigger the formation of clumps or other
features through the PeI. The largest grains in the lowest-gas runs are so
poorly coupled that resolving the PeI growth timescale associated with them
would require prohibitively long run times; in these low-gas runs, any
participation of large grains in PeI-induced effects would presumably require
some complex interplay between small and large grains. For instance, small
grains, even unbound ones, could trigger gas overdensities that yield
substantially shorter values of {\St}, leading to better coupling of larger
grains.

\section{Results} \label{s:results}

\begin{figure}
\resizebox{\columnwidth}{!}{\includegraphics{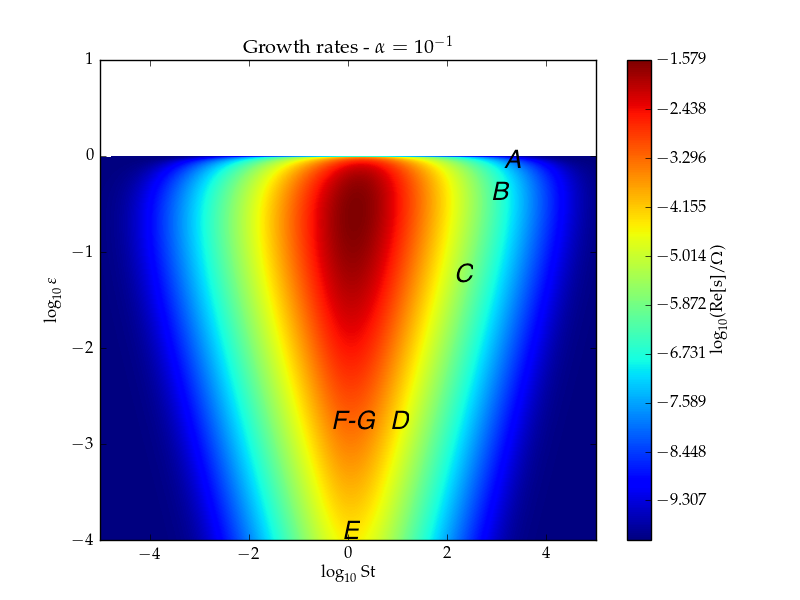}}
\caption[Growth rates as a function of Stokes number]{Maximum growth rates of
  the photoelectric instability as a function of Stokes number and
  dust-to-gas ratio, for a system with turbulent viscosity of
  $\alphaSS=0.1$. The runs of this paper are labeled in the
  plot. The Stokes number chosen is in the middle of the range of the
  actual nonlinear simulations. The actual range of $\St$ of 
  each run should span one order of magnitude in each direction. There
exists no linear instability for dust-to-gas ratio above unity.}
\label{fig:growthrateslabel}
\end{figure}

To guide the interpretation of the simulations, we compute the analytical
growth rates of the PeI as a function of Stokes number for the range
considered. This is done by solving Eq. 26--29 of \citet{LyraKuchner13}.
Without viscosity, the growth rates would grow unboundedly with wavenumber,
eventually getting unphysically high at the grid scale. In reality the growth
rates drop abruptly when the viscous range is approached. Because of this, we
regularize the system with Laplacian viscosity. The result for $\alphaSS = 0.1${, as in run G,}
is plotted in \fig{fig:growthrateslabel}. The figure shows the maximum growth
rate $s$ as a function of Stokes number and dust-to-gas ratio. 
{The labels A--G indicate the minimum Stokes numbers and dust to gas ratios corresponding to simulations A--G (if runs A--F had Laplacian viscosity rather than hyperviscosity).} Above dust-to-gas ratio unity no linear instability exists. The symmetry with respect to $\St=1$, seen in the nonlinear simulations of \citet{LyraKuchner13}
(Supplement Fig. 2 of that paper) is reproduced. The runs of this paper are
labeled in the graph. The $\St$ value chosen as representative of the run is at
$r=1$. In each simulation, the range of $\St$ should reach an order of
magnitude in each direction. 

\begin{figure*}
\centering
\includegraphics[width=0.95\textwidth]{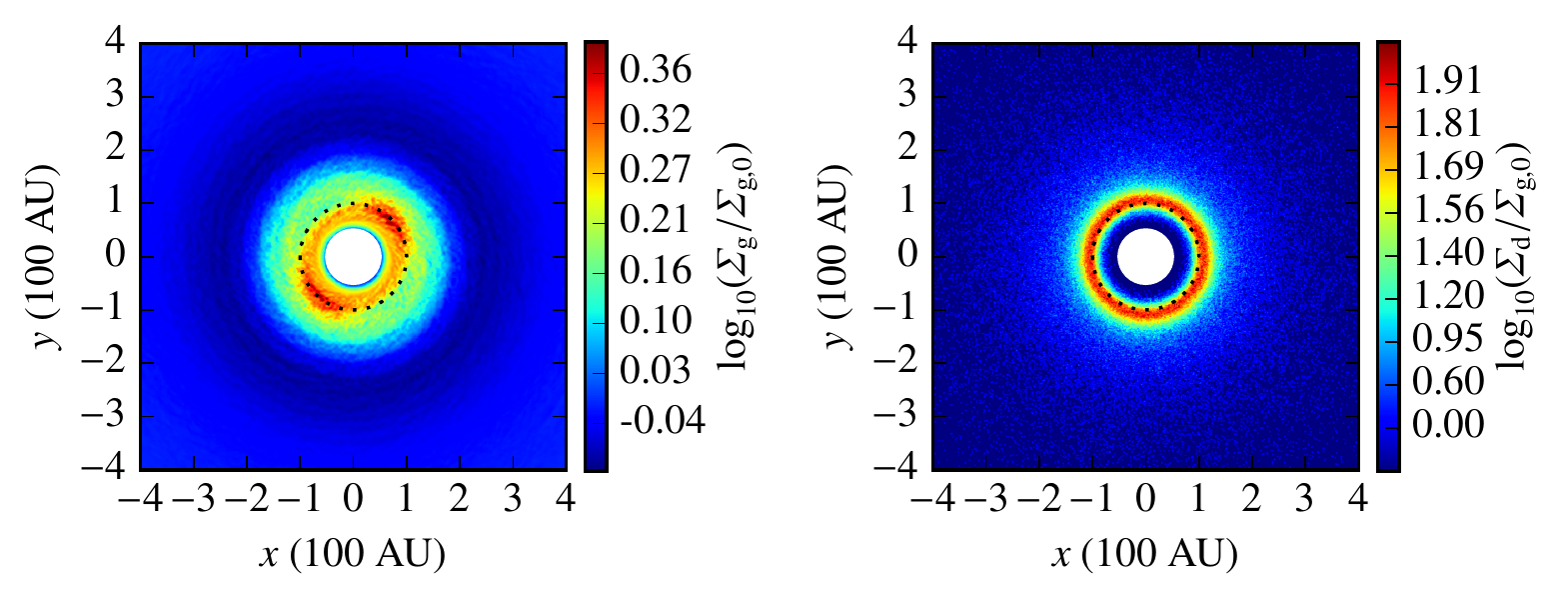}
\caption[Gas and dust surface densities after 400 orbits for run A]{\runcaption{A}{\gasA}{\dustA} The dust (right panel) shows no
indication of perturbation by the gas. The gas vortices (left panel) just
outside the birth ring represent a factor of two overdensity, still too small to
substantially shorten the large values of {\St} for the dust.}
\label{f:runA}
\end{figure*}

\begin{figure*}
\centering
\includegraphics[width=0.95\textwidth]{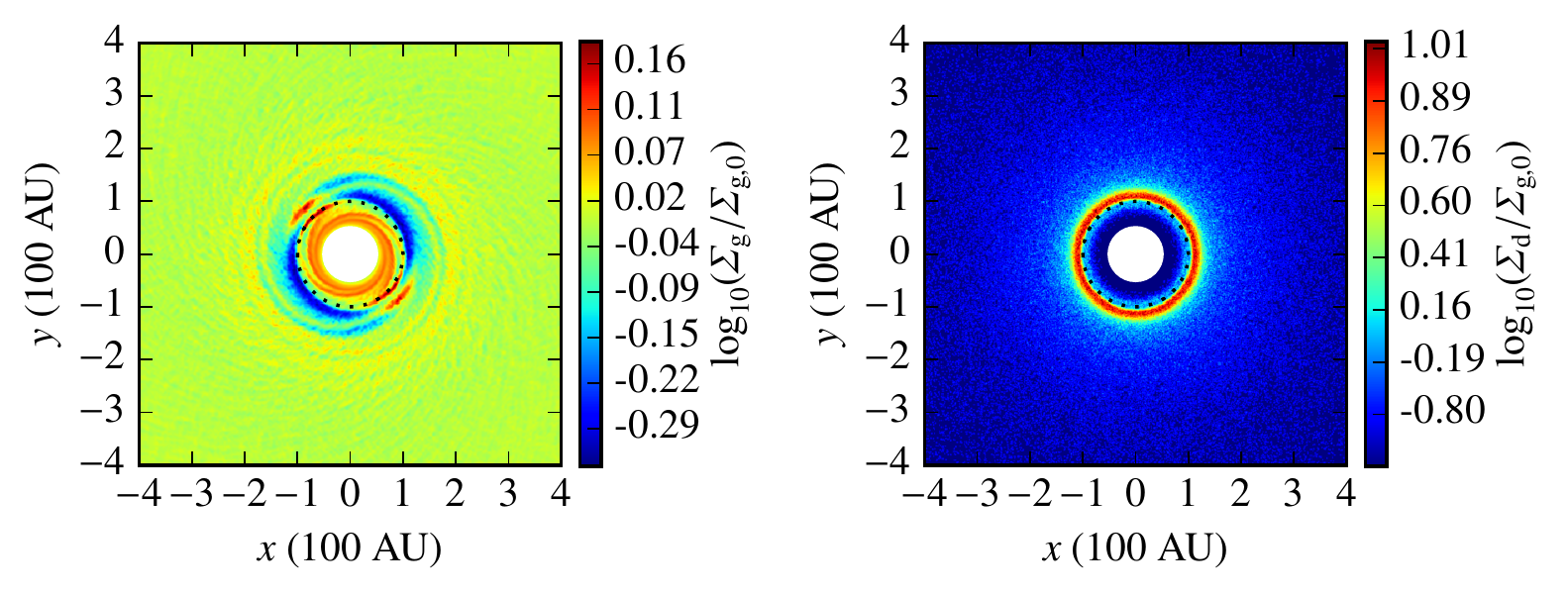}
\caption[Gas and dust surface densities after 400 orbits for run B]{\runcaption{B}{\gasB}{\dustB} Compared with run A (Fig.~\ref{f:runA}),
the dust ring is radially narrower, but still shows no signs of
PeI-induced clumping.}
\label{f:runB}
\end{figure*}

\begin{figure*}
\centering
\includegraphics[width=0.95\textwidth]{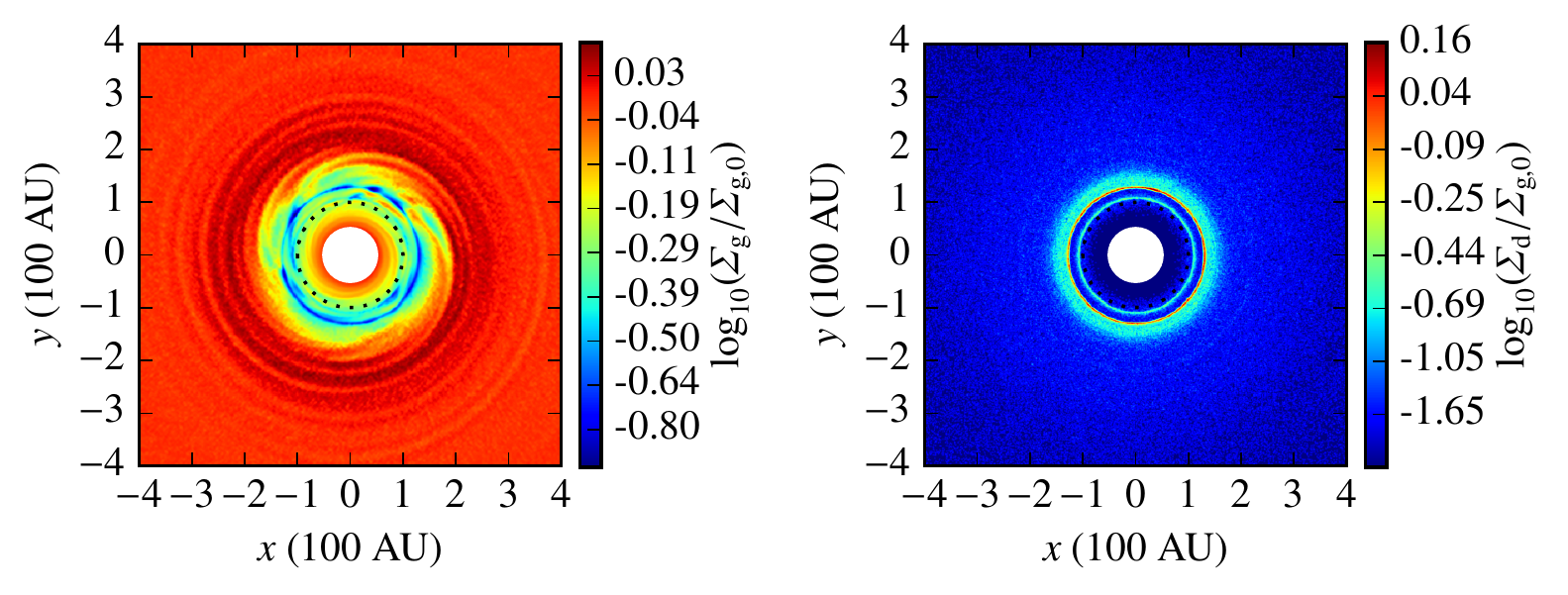}
\caption[Gas and dust surface densities after 400 orbits for run C]{\runcaption{C}{\gasC}{\dustC} The dust distribution shows two
closely spaced concentric rings, consistent with dust clumping due to the PeI.}
\label{f:runC}
\end{figure*}

\begin{figure*}
\centering
\includegraphics[width=0.95\textwidth]{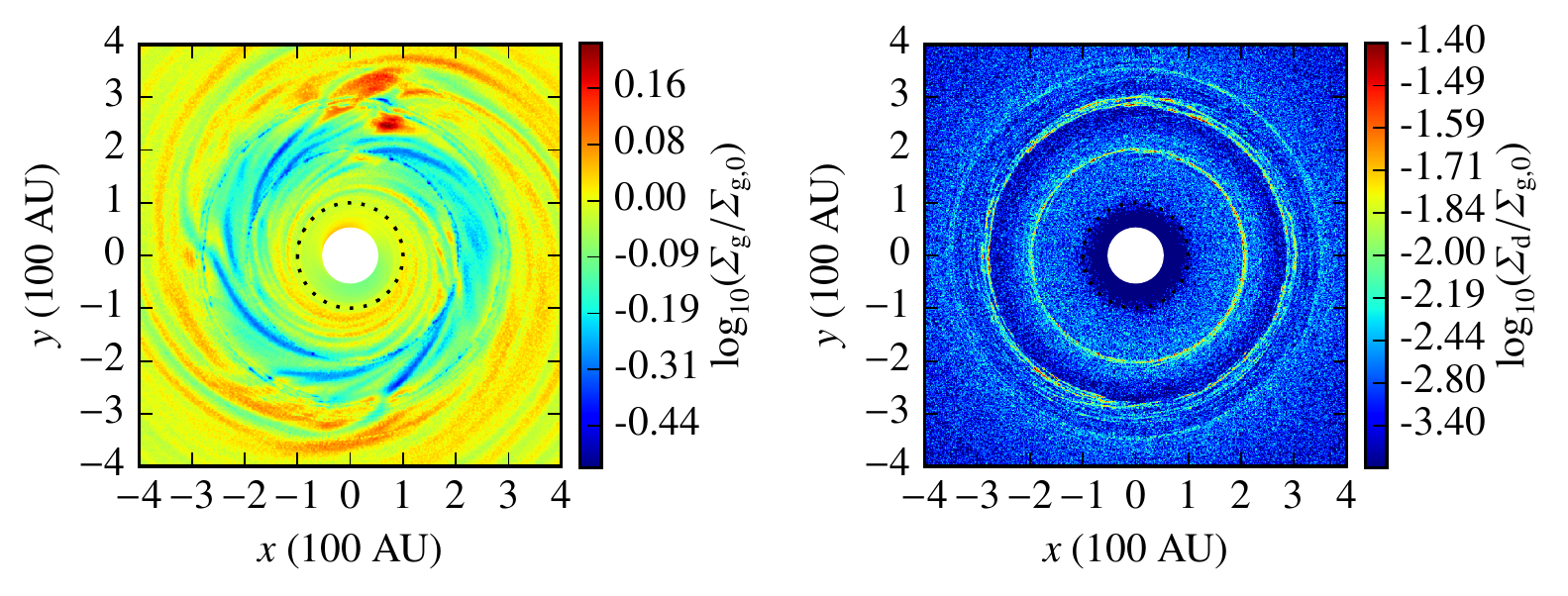}
\caption[Gas and dust surface densities after 400 orbits for run D]{\runcaption{D}{\gasD}{\dustD} The gas distribution (left panel) shows
greater non-axisymmetric structure than in run C (Fig.~\ref{f:runC}). The dust
distribution (right panel) shows two prominent ring structures and a third
fainter one (the outermost). Ripple structures in the dust spaced a few AU apart
also appear near each of the three rings, strongly reminiscent of the
tightly-packed arcs and rings seen in the radiation pressure-free models of
\citet{LyraKuchner13}.}
\label{f:runD}
\end{figure*}

\begin{figure*}
\centering
\includegraphics[width=0.95\textwidth]{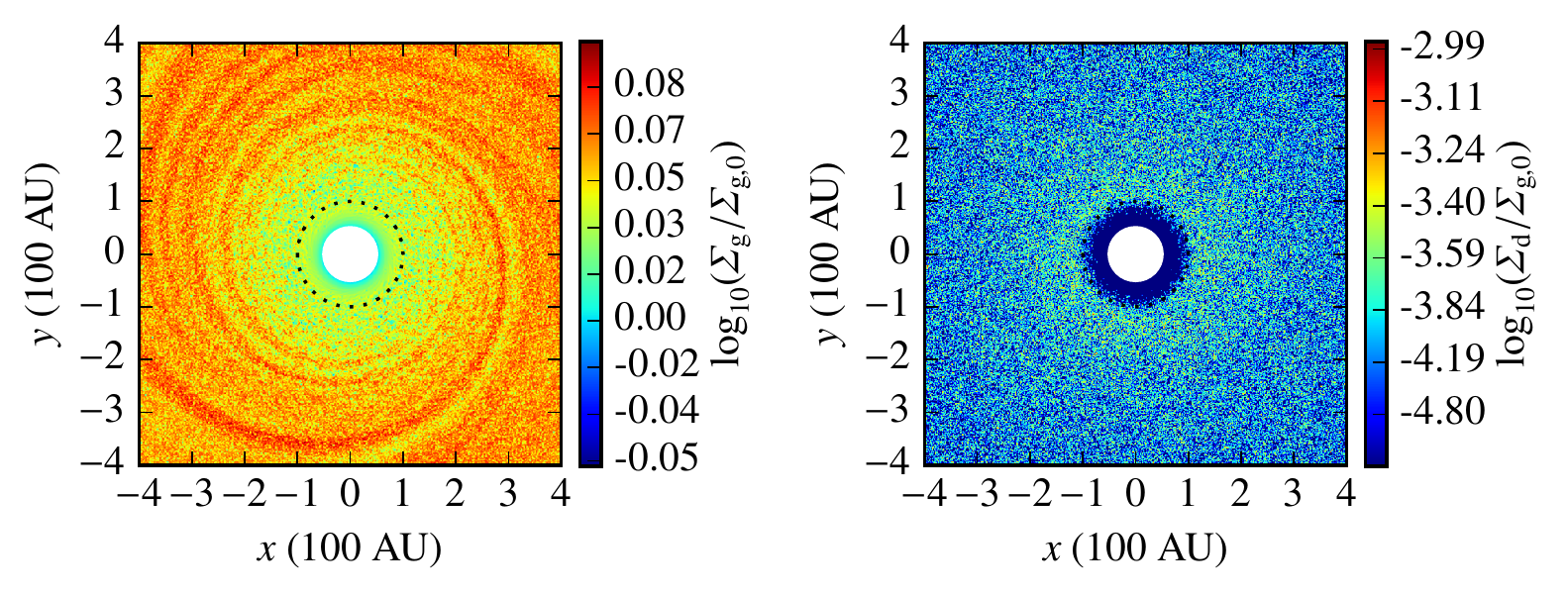}
\caption[Gas and dust surface densities after 400 orbits for run E]{\runcaption{E}{\gasE}{\dustE} The lack of structure in the dust
distribution (right panel) reflects a long PeI growth time for such a low
dust-to-gas ratio.}
\label{f:runE}
\end{figure*}

\begin{figure*}
\centering
\includegraphics[width=0.95\textwidth]{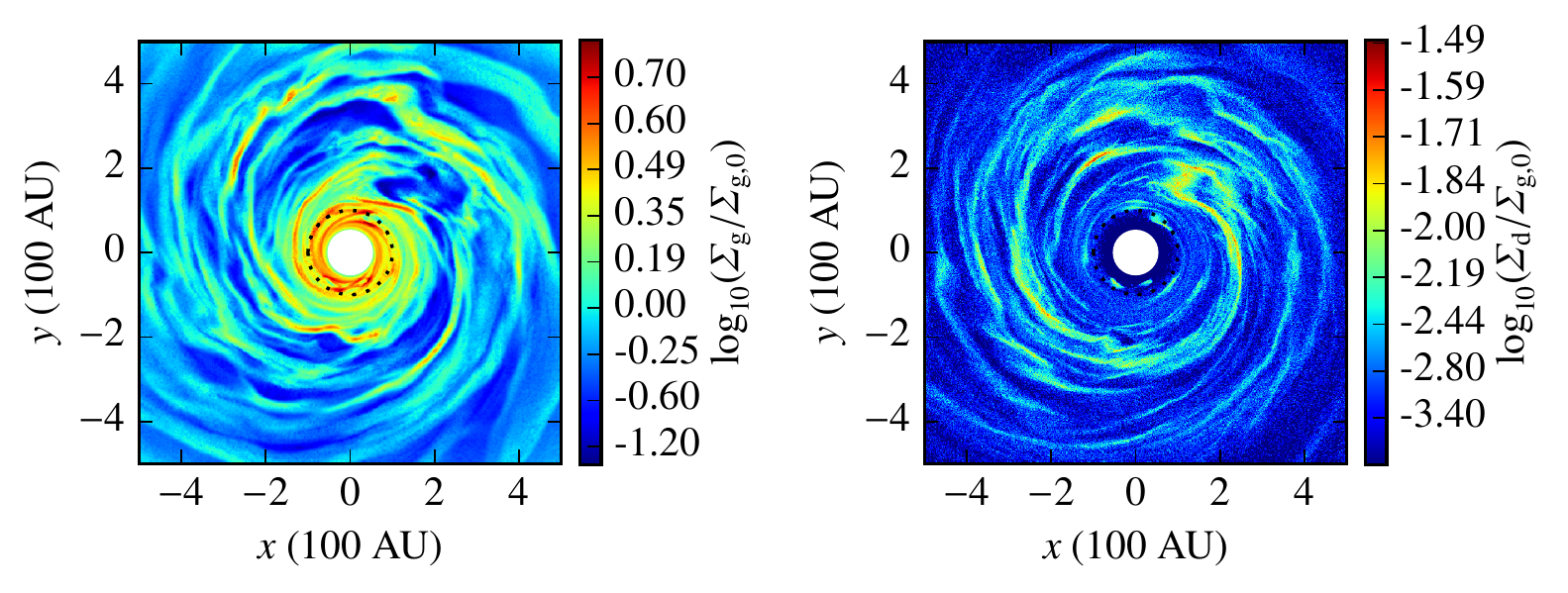}
\caption[Gas and dust surface densities after 400 orbits for run F]{\runcaption{F}{\gasF}{\dustF} Increasing the dust mass by a factor of
ten compared with run E (Fig.~\ref{f:runE}) yields the return of the photoelectric
instability. Compared with runs A-\-E, both the gas and dust (left and right
panels, respectively) follow non-axisymmetric distributions.}
\label{f:runF}
\end{figure*}

\begin{figure*}
\centering
\includegraphics[width=0.95\textwidth]{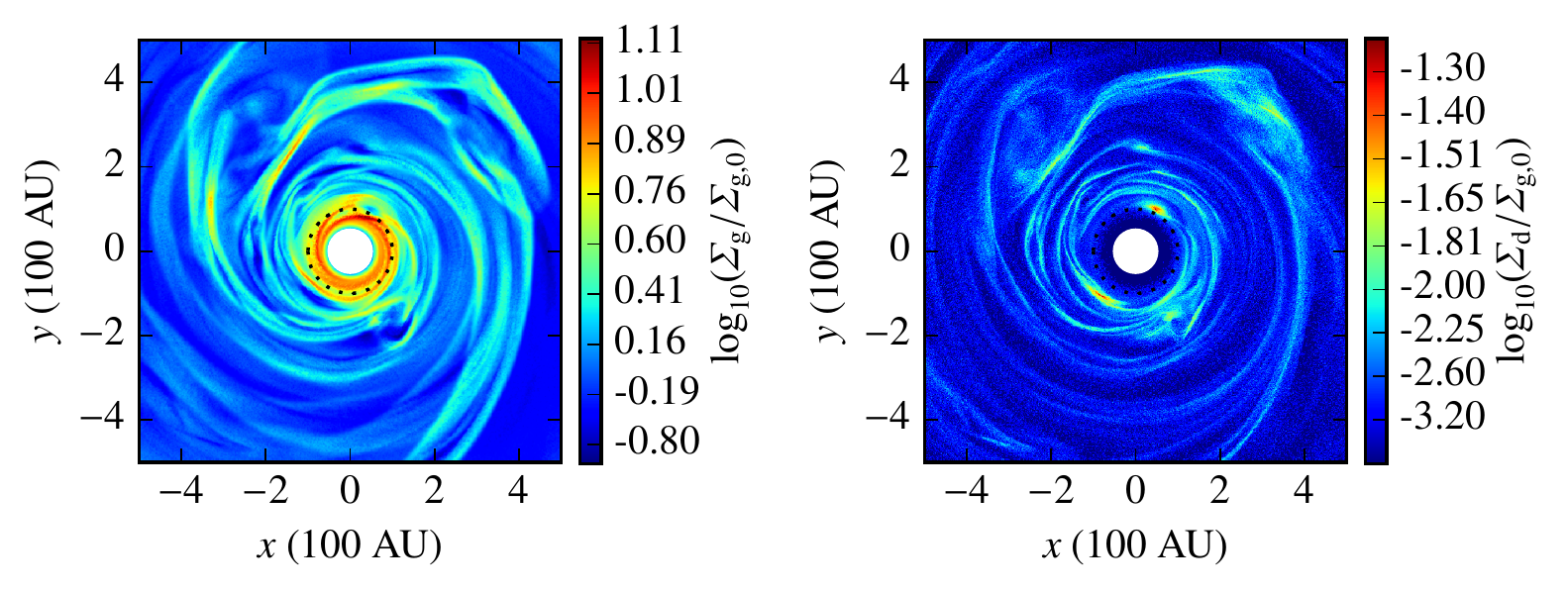}
\caption[Gas and dust surface densities after 400 orbits for run
G]{\runcaption{G}{\gasG}{\dustG;\ \alphaSSref=0.1} The addition of a
substantial low-order viscosity does not impede the onset of the photoelectric
instability, but it smooths out much of the high-frequency structure seen in run F
(Fig.~\ref{f:runF}).}
\label{f:runG}
\end{figure*}

Gas and dust surface densities after 400 orbits at 100~AU for runs A--G are
shown in Figures \ref{f:runA}--\ref{f:runG}. For simplicity, we hereafter use
the term ``orbits" to indicate orbits at the reference radius $\rBR=100~\AU$.

For runs A and B (Figs.~\ref{f:runA} and \ref{f:runB}), we see that a narrow,
axisymmetric dust ring forms just outside the birth ring, with a central
surface density corresponding with $\tau\approx5\times10^{-3}$. In the gas,
two vortices form just outside the birth ring. In run B, a gas gap appears,
along with two additional vortices appear on the opposite side of the gap from
the first two, matching them in azimuth. Each vortex orbits at sub-Keplerian
speed, with an orbital frequency approximately 90\% of that expected from
Keplerian rotation.

\begin{figure*}
\centering
\includegraphics{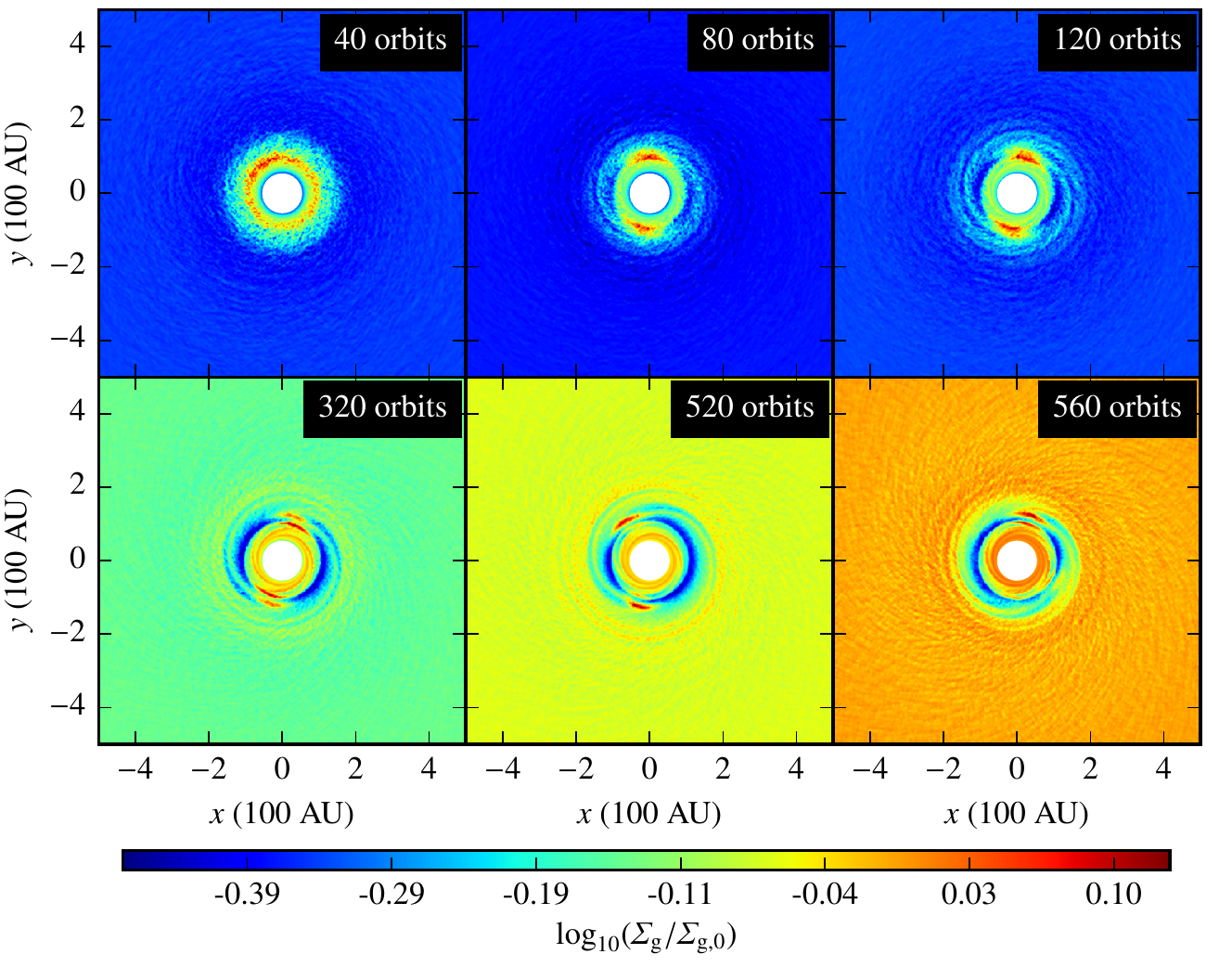}
\caption[Gas surface density over time for run B]{Gas surface density over time for run B. After 560 orbits, only one
vortex remains following a long-lived $m=2$ pattern that emerges after dozens of orbits.}
\label{f:runB_rhoSnaps}
\end{figure*}

In order to further investigate this behavior, we plot the gas surface density
over time for run B in Figure~\ref{f:runB_rhoSnaps}. In the first few dozen
orbits, a single vortex ($m=1$) emerges. In the next several dozen orbits, a gas
gap forms, as well as two vortices ($m=2$) just outside it, positioned
$180^\circ$ apart from each other in azimuth; the contemporaneity of the
formation of the gap and the two inner vortices suggests that the Rossby wave
instability \citep{Lovelace1999} may be responsible. For the next several
hundred orbits, two additional ``matching" vortices appear just within the orbit
of the inner gap edge, keeping pace with the outer vortices. After 520 orbits,
the outer vortices have begun to migrate and the inner vortices are much less
pronounced, and after 560 orbits only one inner--outer vortex pair remains.

In runs A--C (Figs.~\ref{f:runA}--\ref{f:runC}), we see that the increasing
value of {\Sigmagasref} (thus reducing {\St}) results in a narrow gas gap
whose depth increases with {\Sigmagasref}. In run D (Fig.~\ref{f:runD}), we see
a shallower, wider gap threaded by several spiral arm structures (not seen in
the models of \citeauthor{LyraKuchner13} \citeyear{LyraKuchner13}).

In runs C and D (Figs.~\ref{f:runC} and \ref{f:runD}), compared with runs A and
B, the gas distributions are less smooth and contain non-axisymmetric clumps
and arcs. The dust distributions show concentric rings with factor of 10--20
dust enhancements, reminiscent of the models of \citet{LyraKuchner13}, though
there a fewer rings in runs C and D, which have two and three main ring
structures, respectively. The central optical depths of these rings are of
order $10^{-3}$, with the exception of the outermost ring in run D
(Fig.~\ref{f:runD}), which is about an order of magnitude fainter. In run C,
the anticorrelation of gas and dust just outside the birth ring is reminiscent
of the dust--gas anticorrelation seen in the models of \citet{LyraKuchner13}.

In run D, the dust rings are accompanied by more high-frequency structure,
including some localized ripple patterns (a few AU in scale) strongly
reminiscent of the tightly packed arcs and rings seen in the
\citet{LyraKuchner13} models. These rings are long-lived, having begun to form
after only a few tens of orbits.  {Though the gas responds to the dust on dynamical timescales due to photoelectric heating, in runs A--D the Stokes numbers are high enough that the bound grains ($\betaPR < 1/2$) respond to the gas only on much longer timescales; grains on low-eccentricity orbits see the azimuthally-averaged gas distribution over the course of many orbits, not the gas's spiral structure. This asymmetrical coupling promotes the formation of dust rings. But note that some of the fine structure in the rings, like the bifurcation at the one o'clock position, seems to correspond with clumps in the gas.}

No indications of the photoelectric instability emerge after 400 orbits for run
E (Fig.~\ref{f:runE}). As shown in \fig{fig:growthrateslabel}, the growth rate
for run E is centered at the maximum value of $s\approx 10^{-5}\Omegagas$. After
400 orbits, this low growth rate amounts to a mere 2\% amplification. Runs D
and C, though centered at the same low level of growth rate, and B, centered at
even lower, reach to the left of the diagram and into regions of higher growth
rate, as much as $10^{-2}\Omegagas$, (million-fold amplification in $\approx$ 220
orbits). Run A, though also centered at a low value of growth rate, and not
reaching too deep into regions of high growth rate, may be non-linearly
amplified due to the high dust-to-gas ratio.

For run F (Fig.~\ref{f:runF}), we find that increasing the total dust mass
compared with run E yields the return of the photoelectric instability. The
photoelectric instability radically transforms both the gas and dust
distributions; the dust clumps and arcs seen in the right panel of
Figure~\ref{f:runF} correspond with $\tau \approx 10^{-2}$, representing
factor of 5--10 enhancement over the local dust surface density.

\begin{figure*}
\centering
\includegraphics[width=0.95\textwidth]{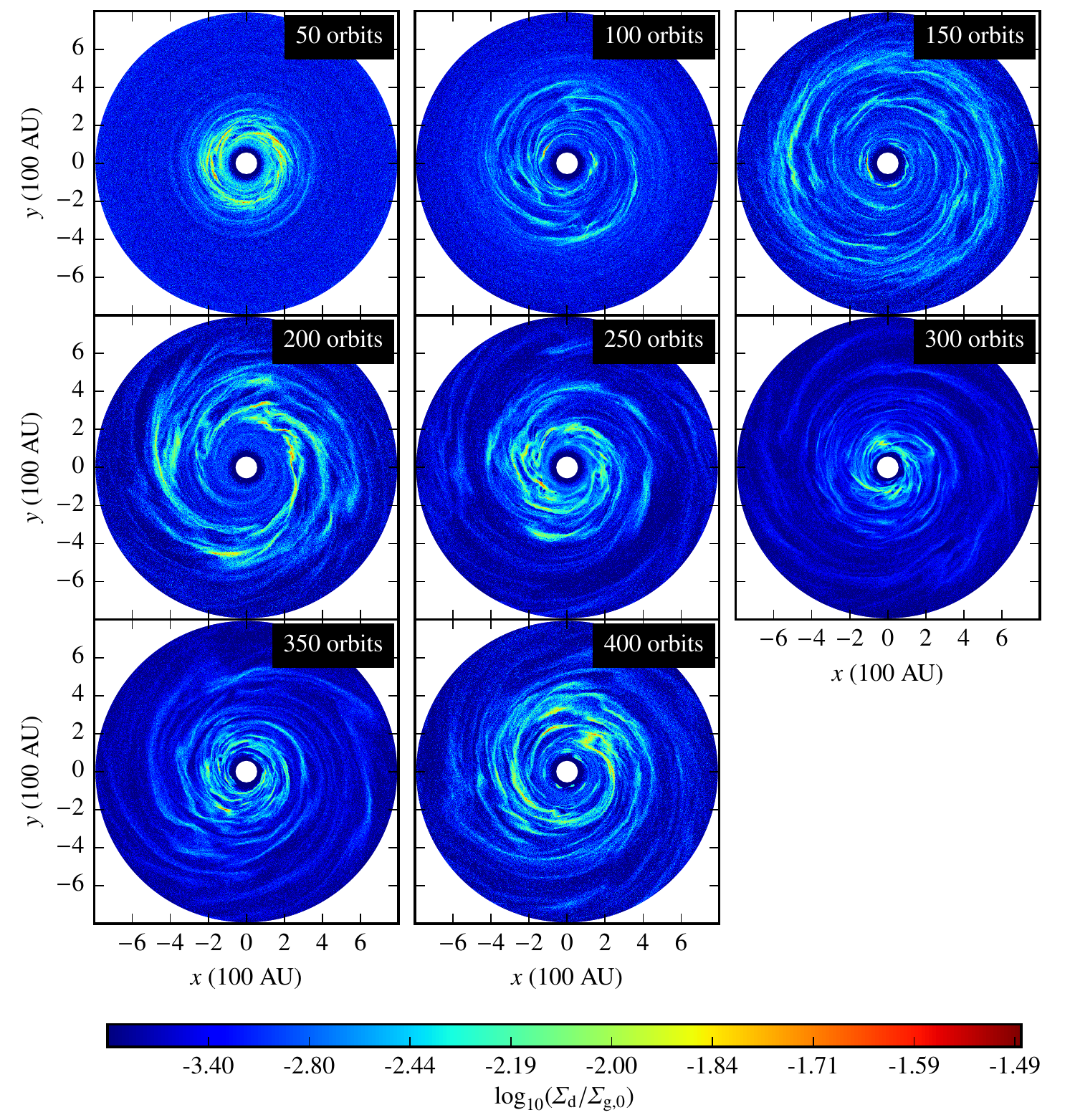}
\caption[Dust surface density over time for run F]{Dust surface density over time for run F. The photoelectric
instability arises quickly, continuously redistributing dust into
non-axisymmetric structures throughout the disk over the full course of the
simulation. Though the radial distribution of the dust varies nonsecularly over
time, the two-dimensional dust structure is consistently dominated by clumps
and arcs that form and dissipate on short timescales $\approx1-10$~orbits.}
\label{f:runF_rhopSnaps}
\end{figure*}

In Figure~\ref{f:runF_rhopSnaps}, we show the dust surface density every 50 orbits
for 400 orbits. We find that the photoelectric instability sets in quickly, but
has no obvious secular effect on the global structure of the disk. The behavior
is dominated by transient clumps and arcs that appear and disappear on
timescales of orbits to dozens of orbits. {Qualitatively, the behavior of run F 
is quite different from that of run D, where the dust remains in mostly smooth rings. As seen in Figure~\ref{f:tauvsbeta}, for run D, bound dust grains are weakly coupled with the gas, generating rings as discussed above. In run F, a large range of grain sizes are both bound and well-coupled to the gas. Setting ${\St}=1$ and ${\beta}=1/2$ in Equation~\ref{e:tauvsbeta}, we find that the threshold for having grains that are both bound and well-coupled occurs at a gas surface density of
\begin{equation} \label{e:threshold}
{\Sigmagas} > {1.8 \times10^{-4}~{\gram}\,{\cm}^{-2}} \left(\frac{\Lsun}{\Lstar}\right)  \ \left(\frac {\Mstar}{\Msun}\right).
\end{equation}
}

The gas and dust distributions for run F (Fig.~\ref{f:runF}) are strikingly
similar to the results of the \citet{LyraKuchner13} model that excludes drag
backreaction on the gas. In the absence of radiation pressure, where both the
gas and dust move on roughly circular orbits, the expansion of gas due to high
pressures induced by photoelectric heating will undergo Coriolis rotation. In
models with backreaction, this rotation is opposed by the backreaction from the
dust, and axisymmetry of the system is maintained. It is possible that when a
dust grain is placed on a highly eccentric orbit by radiation pressure, it heats
gas in a given region, creating a pressure maximum, but does not linger in a
nearby circular orbit where it can stabilize the gas through drag backreaction.

\begin{figure}
\centering
\includegraphics[width=3.25in]{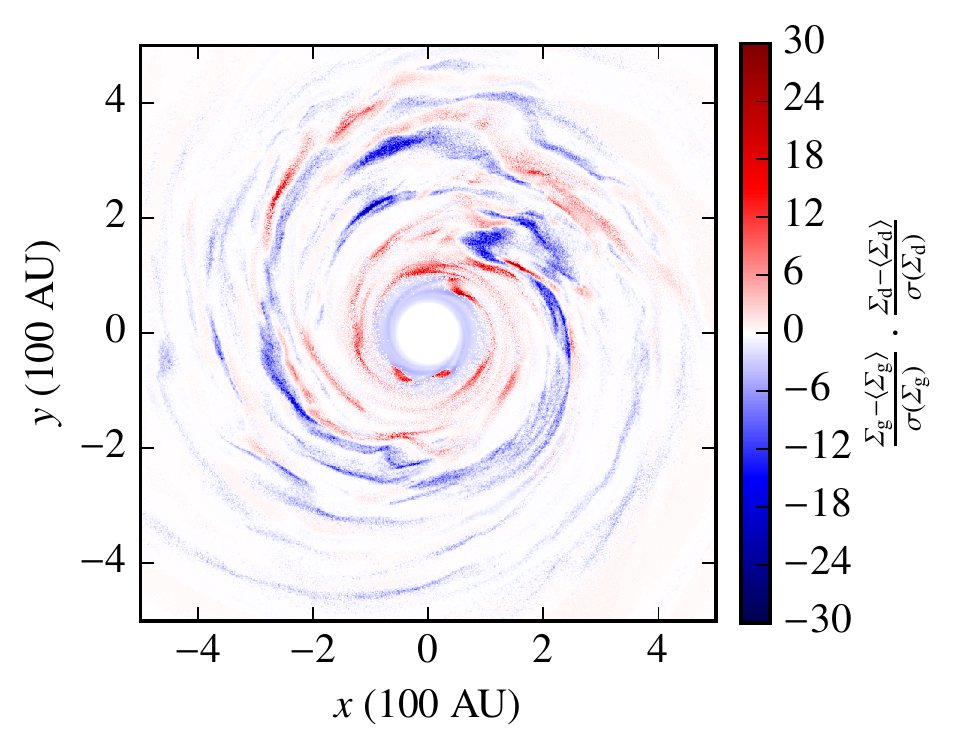}
\caption[Gas--dust correlation map for run F]{Gas--dust correlation map for run
F ($\Sigmagasref=\gasF$; $\Mdust=\dustF$). The gas and dust correlate (red
regions) and anticorrelate (blue regions) in equal measure, contrary to the
alternating (i.e., anticorrelating) patterns of gas and dust seen in the models
of \citet{LyraKuchner13}.}
\label{f:runF_gdcorr}
\end{figure}

\citet{LyraKuchner13} find that in the presence of drag with backreaction and
photoelectric heating, the gas and dust mutually displace each other, leading to
alternating rings of gas and dust throughout the disk. It is apparent in
Figures~\ref{f:runA}--\ref{f:runC} that when a small amount of gas is present,
dust displaces it, creating a gas gap. In order to test whether gas and dust
anticorrelate when larger quantities of gas are present, in
Figure~\ref{f:runF_gdcorr} we plot the product of {\Sigmagas} and {\Sigmadust}
(specifically, their mean-subtracted and standard deviation-normalized values)
for run F after 400 orbits. We find that the dust and gas correlate (red
regions) and anticorrelate (blue region), in roughly equal measure.

We find that the inclusion of a very strong low-order gas viscosity term in run
G (Fig.~\ref{f:runG}) yields results are fairly similar to the no-viscosity case
(run F; Fig.~\ref{f:runF}). It does, however, lead to the smoothing of some of
the small-scale structure in both the gas and dust distributions seen in run F;
dust surface density enhancement are of order factor of 5, somewhat smaller than
seen in run F. The small, transient dust clumps seen in run F are less numerous
and less elongated in run G, but are nonetheless present, also appearing and
disappearing over orbital timescales. This result is consistent with the
predictions and models of \citet{LyraKuchner13}, where viscosity suppresses the PeI
at high wavenumbers, smoothing small-scale structure but not impeding the
ability of the PeI to substantially reshape the disk.

\begin{figure}
\centering
\includegraphics{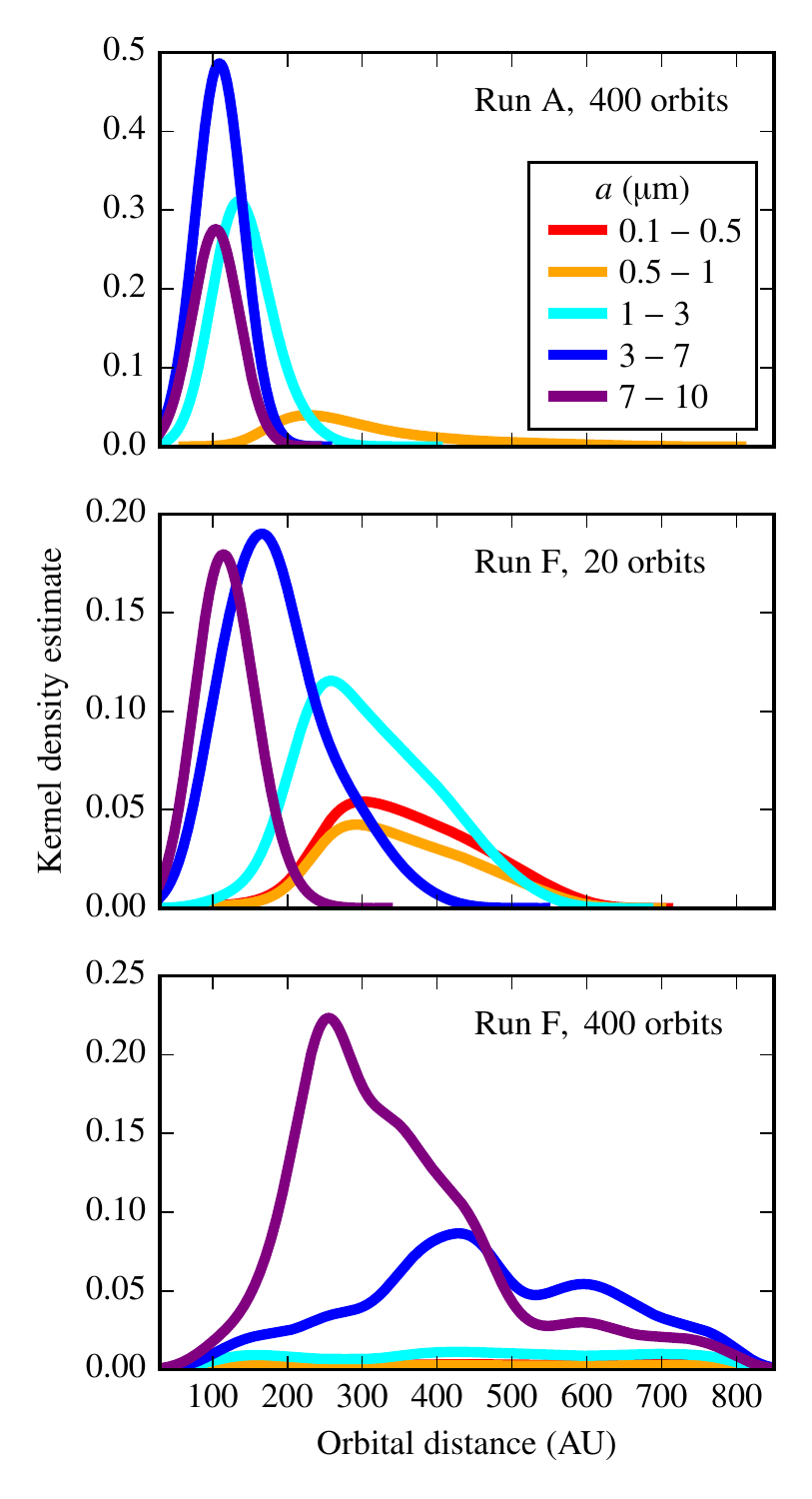}
\caption[Radial dust distributions by grain size]{Radial distributions for five grain size bins for run A after 400
orbits (upper panel), run F after 20 orbits (middle panel), and run F after 400
orbits (lower panel). When little gas is present (run A, upper panel), large,
bound grains remain on circular orbits near the birth ring, almost completed
unperturbed by the sparse gas. At higher gas densities (run F), small grains
trigger the photoelectric instability outside the birth ring (middle panel),
but the resulting gas flows also entrain larger grains, carrying them away from
the birth ring, eventually dominating nearly the full radial extent of the disk
(lower panel).}
\label{f:r_hist}
\end{figure}

In order to explore the roles played by different grain sizes in creating and
sustaining the structures seen in the right-hand panels of
Figures~\ref{f:runA}--\ref{f:runG}, we examine the radial distributions of dust
in several bins of grain size, choosing runs A and F as representative cases.
The upper, middle, and lower panels of Figure~\ref{f:r_hist} show radial dust
distributions for run A after 400 orbits, run F after 20 orbits, and run F after
400 orbits, respectively.

The radial dust distributions shown in the upper panel of Figure~\ref{f:r_hist}
confirm that it is large particles (i.e., small $\beta$, corresponding with
nearly circular orbits) that constitute the dust ring seen in run A
(Fig.~\ref{f:runA}). This is consistent with the fact that these sharp rings
take dozens of orbits to form, reflecting the large values of {\St} for
large grains. The small values of $\beta$ place these large grains on circular
orbit, keeping them close to the birth ring.

In the middle and lower panels of Figure~\ref{f:r_hist}, we identify two
important grain size-dependent effects. The first, seen in the lower panel, is
that even large grains (7--10~{\micrometer}) migrate outward from their
low-eccentricity orbits in and around the birth ring, having now had enough time
($t>\St\cdot\OmegaK$) to be entrained in PeI-induced flows. The second is that
smaller grains are more efficiently entrained by the denser gas, leading
ordinarily unbound particles ($\beta<1/2$) to remain in the disk.
The middle panel shows the grain distribution for run F after only 20 orbits, at
which point the photoelectric instability is just forming. Small-to-medium
grains (0.1--3~\micrometer), as they are being blown outward by radiation
pressure, accumulate outside the birth ring and trigger the photoelectric
instability. Meanwhile, larger and therefore more weakly coupled grains migrate
outward by gas drag over the course of orbits to tens of orbits. The dominance
of large grains seen in the lower panel of Figure~\ref{f:r_hist} suggests that
while small, unbound grains help to trigger the photoelectric instability and
participate in the resulting transient structures, many of them will ultimately
be blown out, and the dust structure seen in Figure~\ref{f:runF} eventually
becomes dominated by larger, bound grains. This suggests that even in disks
around more massive (say, A-type) stars, the extreme radiation pressure on dust
grains (and subsequent lack of bound grains) does not necessarily inhibit the
formation of the photoelectric instability.
{Note that Figures 3--9 show the dust surface density, which tends to be dominated by larger grains. But we found that plotting the optical depth (not shown) results in images that are nearly indistinguishable, despite emphasizing smaller grains.}

\section{Conclusions} \label{s:conclusions}

We have produced hydrodynamical models of optically thin disks with gas and
dust, simultaneously incorporating photoelectric heating and stellar radiation
pressure on dust grains. We find the following:

\begin{itemize}
\item
The emergence of the photoelectric
instability \citep{Klahr2005, Besla2007, LyraKuchner13} is not impeded by the radiation pressure associated with a solar-type star.

\item
The PEI growth rate is small for gas surface densities $ \Sigmagas < 10^{-6}~\gcmtwo$, but fast enough at higher gas surface densities that we see the PEI create dust density enhancements of up to a factor of 20 in our runs of just 400 orbits. 

\item
For a modest level of gas ($10^{-6}~\gcmtwo<\Sigmagas<10^{-5}~\gcmtwo$; runs C
and D), the photoelectric instability gives rise to axisymmetric dust rings over
the course of dozens of orbits, as well as azimuthal structure in the gas. The
dust and gas show the anticorrelation bevahior predicted by \citet{LyraKuchner13}.

\item
For a higher level of gas ($\Sigmagas=10^{-4}~\gcmtwo$; run E), and similar dust-to-gas ratio {\epsdtog}, (runs F and G), the PeI emerges over the course of dozens of orbits, leading to
erratic spiral structure throughout the disk (resembling the backreaction-free
model of \citeauthor{LyraKuchner13} \citeyear{LyraKuchner13}), with small-scale structure
appearing and disappearing over the course of orbits. In these models, there is
no overall tendency of the dust and gas to anticorrelate. 

\end{itemize}

The value of $\Sigmagas$ required to generate the PeI
on short timescales in a given system, however, will depend on many parameters,
including spectral type, total dust mass, initial gas profile, and grain size
range.  In some of our models with higher level of gas ($\Sigmagas=10^{-4}~\gcmtwo$; run E), but low
effective dust-to-gas ratio {\epsdtog}, we found the PeI growth rate too slow to be captured by our simulations.

Previous models of debris disks with gas have lacked the physics to capture
hydrodynamical instabilities like the PeI. \citet{Thebault2005} and
\citet{Krivov2009} present models of debris disks with gas spanning a range of
gas surface densities similar to the range used in the current work (though
other key parameters vary between these works, such as stellar spectral type
and the initial radial profile of the gas). In both cases, dust grains
generated in a birth ring experience aerodynamic drag with a static gas cloud,
precluding the emergence of the PeI, which requires dust--gas heating and drag
backreaction. These one-dimensional models yield dust density distributions
that decrease smoothly and monotonically with orbital radius, making them
readily distinguishable from the clumps, arcs, and narrow rings seen in the
models presented in Section~\ref{s:results}.

Other models of optically thin disks reveal more complex morphologies that are
not as readily distinguished from the results presented in this paper. For
instance, the models of cataclysmic massive body collisions produced by
\citet{Kral2015} produce non-axisymmetric structures that could be difficult to
distinguish from the dust distributions shown in Figures~\ref{f:runF} and
\ref{f:runG}, especially for disks in an edge-on viewing configuration.
Nonetheless, the models of \citet{Kral2015} show smoother, more organized
structure compared with the higher-frequency, more erratic structure seen in
runs F and G (Figs.~\ref{f:runF} and \ref{f:runG}), and also do not show
concentric, axisymmetric rings as in runs C and D. Comparisons of dust
distributions at multiple wavelengths for a given disk could also help to
disentangle these two effects, given that differential behavior by grain size
should be greater for aerodynamic effects than gravitational ones.

\citet{Augereau2004} model a circumstellar debris-only disk with an external
stellar perturber in order to study the origins of the spiral morphology of the
HD~141569 disk. In general, their models produce smooth spiral structure in the
circumprimary disk, however for perturbers with eccentric orbits, the structure
produced in the disk is less smooth and somewhat reminiscent of the dust
distributions seen in runs F and G (Figs.~\ref{f:runF} and \ref{f:runG}). Here
again, multiwavelength image comparisons may be necessary to distinguish these
models. Given that none of the numerical models so far produced of the PeI have
given rise to smooth spiral arms of dust (\citeauthor{LyraKuchner13}
\citeyear{LyraKuchner13} and the current work), it would seem that perturbation by a
massive companion is currently a more plausible explanation for such structures
(though this may change as the PeI is modeled throughout a larger parameter
space). 

For a number of other observed disk morphologies, however, the PeI may provide a
more plausible explanation than the presence of a massive perturber. The
resemblance of the dust distributions for runs C and D (Figs.~\ref{f:runC} and
\ref{f:runD}) to the concentric rings seen in the scattered light profile of the
disk around HIP~73145 \citep[spectral type A;][]{Feldt2017} is particularly
striking. The large grains are relatively compactly distributed, while small
grains experience considerable radiation pressure. Though we do not model A-type
stars in the current work, our models suggest that the photoelectric instability
provides a promising explanation for such features, which can be triggered by
unbound grains.

Also, the non-axisymmetric clumps and spiral arms seen in the gas distributions
in several of our models suggest that interactions between gas and
small-to-medium grains could underlie the asymmetric structures seen in the
49~Ceti and AU~Mic edge-on disks \citep{Hughes2017, Boccaletti2015}, though
explanations for the AU~Mic moving blobs involving only rocky body collisions
(no gas) have also been proposed \citep{Sezestre2017, Chiang2017}. We also
underscore that so far the PeI is the best candidate for producing arcs, a
feature that is not predicted from planet-disk interaction. That the PeI leads
to arcs was predicted by \citet{LyraKuchner13}, before the discovery of these
features in the disk around HD~141569A \citep{Perrot2016}.

The broad resemblance of our models to several observed systems notwithstanding,
models that include more physical detail---magnetic fields, multiple gas
species, and so on---and also explore a wider range of parameters (especially
stellar spectral type) will help to confirm that the photoelectric instability
can indeed provide a plausible explanation for these diverse and intriguing disk
morphologies.

Future hydrodynamical models of optically thin disks should explore a number of
physical processes not explored in the current work: 

\begin{enumerate}

\item The magnetorotational instability (MRI) may operate efficiently in debris
disks \citep{Kral2016mri}. The inclusion of a low-order viscous term in run G
produced an observably different dust distribution from run F; the role of
MRI-induced turbulence in redistributing gas, and subsequently dust, should be
explored in detail.

\item Future studies should explore the competing roles of gas--gas and
dust--gas photoelectric heating to determine the precise realm of disk parameter
space in which dust--gas photoelectric heating is dynamically important.
\citet{Kral2016} point out that in a carbon-rich disk like that around
$\beta$~Pic, heating by photoelectrons released from carbon may overwhelm the
heating produced by those released from dust. This in turn may inhibit the
photoelectric instability by inhibiting the formation of pressure maxima.

\item In the current work, we model only a single gas species.  However, as
noted by \citet{Xie2013}, different gas species can experience different values
of $\beta$. The presence of a modest radiation force on a certain gas species
with, say, $\beta=0.2$ would significantly affect the mutual velocity of that
species with the dust, potentially altering the effects of the PeI.

\item Future investigations should explore the role of Poynting--Robertson drag
in the context of the photoelectric instability. The findings of
\citet{LyraKuchner13} suggest that even large, poorly coupled grains will eventually
give rise to the photoelectric instability, but on very long timescales, inward
drift due to PR drag could inhibit or affect this process.

\item In the current work, we have only modeled disks around solar-type stars.
Around A-type stars, where many debris disks, including those with unusual
morphologies, are found, the radiation force on dust grains will be considerably
greater. This means that the bound grains will be larger (though the results
presented in Fig.~\ref{f:r_hist} suggest that even unbound grains---small and
well-coupled---can trigger the PeI on their way out of the disk). Large and
therefore weakly-coupled grains can generate and participate in the
photoelectric instability, however the resulting structures will take longer to
emerge (a likely explanation for the lack of dust clumping in runs A and B). It
is possible for instance that the spatial and temporal frequencies of any
non-axisymmetric structures resulting from gas--dust interactions involving
large grains will be larger due to the larger values of {\St}.

\end{enumerate}

Once the interactions of dust, gas, and radiation in optically thin disks are
better understood, the signposts of embedded planets can be more accurately
modeled. The substantial effect of drag backreaction and photoelectric heating
seen in the planet-free models of the current work suggests that planet--disk
interactions in optically thin disks may manifest themselves very differently
than they do in existing, simpler models. The formation of gaps, rings, and
other disk morphologies associated with planets may be inhibited, enhanced, or
otherwise affected by the presence of gas.

\acknowledgments

{We thank Yanqin Wu and Ruobing Dong for commenting on an early version of this manuscript.} W. L. acknowledges support of Space Telescope Science Institute
through grant HST Cycle 24 AR-14572 and the NASA Exoplanet Research
Program through grant 16- XRP16 2-0065. M.K. acknowledges support
provided by NASA through a grant from the Space Telescope Science
Institute (HST Cycle 21 AR-13257.01), which is operated by the
Association of Universities for Research in Astronomy, Inc., under
NASA contract NAS 5-26555.

\bibliography{bib.bib} 

\end{document}